\documentclass[twocolumn]{svjour3}          
\smartqed  
\usepackage[utf8]{inputenc}
\usepackage{graphicx,bm,amssymb,amsmath}
\usepackage[usenames]{color}

\usepackage{ulem} 

\newcommand{\ba}{{\bf a}}

\newcommand{\ff}{{\bf f}}
\newcommand{\bg}{{\bf g}}

\newcommand{\bl}{{\bf l}}

\newcommand{\br}{{\bf r}}

\newcommand{\bv}{{\bf v}}

\newcommand{\bx}{{\bf x}}
\newcommand{\by}{{\bf y}}
\newcommand{\bz}{{\bf z}}

\newcommand{\bF}{{\bf F}}
\newcommand{\bG}{{\bf G}}

\newcommand{\bT}{{\bf T}}

\journalname{Nonlinear Dynamics}
\begin{document}

\title{The theory of the chain fountain revisited}



\author{Drago\c{s}-Victor Anghel}
\institute{D. V. Anghel \at
    Institutul Na\c tional de Cercetare-Dezvoltare pentru Fizic\u a \c si Inginerie \\
    Nuclear\u a -- Horia Hulubei, M\u agurele, Rom\^ania \\
    Tel.: +40746262943\\
    \email{dragos@theory.nipne.ro}
}

\date{Received: date / Accepted: date}

\maketitle

\begin{abstract}
We analyze the chain fountain effect--the chain siphoning when falling from a container onto the floor. We argue that the main reason for this effect is the inertia of the chain, whereas the momentum received by the beads of the chain from the bottom of the container (typically called ``kicks'') plays no significant role.
The inertia of the chain leads to an effect similar to pulling the chain over a pulley placed up in the air, above the container.
In the model used before by the majority of researchers (the so called ``scientific consensus''), it was assumed that up to half of the mechanical work done by the tension in the chain may be wasted when transformed into kinetic energy during the pickup process.
This prevented the chain to rise unless the energy transfer in the pickup process is improved by ``kicks'' from the bottom of the container.
Here we show that the ``kicks'' are unnecessary and both, energy and momentum are conserved--as they should be, in the absence of dissipation--if one properly considers the tension and the movement of the chain.
By doing so, we conclude that the velocity acquired by the chain is high enough to produce the fountain effect.
Simple experiments validate our model and certain configurations produce the highest chain fountain, although ``kicks'' are impossible.
\end{abstract}

\section{Introduction} \label{sec_intro}

The chain fountain~\cite{PhysRevLett.109.134301.2012.Hanna}, also called the Mould effect~\cite{FountainChain.2013.Mould}, has attracted quite some interest in the past few years (Mould's youtube videos on the \textit{siphoning beads} has attracted millions of views).
When a chain falls from a container (for example, a beaker) over its rim, onto the floor, sometimes forms a high arch in the air, which is called the \textit{chain fountain}.
This effect has been investigated both, theoretically and experimentally (see, for example, Refs.~\cite{ProcRSocA.470.20130689.2014.Biggins,EPL.106.44001.2014.Biggins,PhysEduc.50.564.2015.Andrew,ProcRSocA.471.20140657.2015.Virga,AmJPhys.85.414.2017.Pantaleone,ArchApplMech.87.1411.2017.Pfeiffer,FrontiersinPhysics.6.84.2018.Flekkoy,FrontiersinPhysics.7.187.2019.Flekkoy,arXiv1810.13008.Yokoyama} and references therein), and although it is hard to isolate and observe the main physical phenomenon which produces it, it is quite generally (but not unanimously~\cite{arXiv1810.13008.Yokoyama}) accepted that the reaction from the bottom of the container, which ``kicks'' up the beads of the chain as they start flying, is the culprit~\cite{ProcRSocA.470.20130689.2014.Biggins,EPL.106.44001.2014.Biggins,PhysEduc.50.564.2015.Andrew,ProcRSocA.471.20140657.2015.Virga,AmJPhys.85.414.2017.Pantaleone,ArchApplMech.87.1411.2017.Pfeiffer,FrontiersinPhysics.6.84.2018.Flekkoy,FrontiersinPhysics.7.187.2019.Flekkoy}.
As a consequence of this belief, one would expect that the more flexible the chain is, the less likely it is to siphon from the beaker.
This is apparently supported by experiments showing ropes or chains of loosely connected beads crawling over the rim of the container and falling on the other side, onto the floor, pulled by their own weights~\cite{ProcRSocA.470.20130689.2014.Biggins,AmJPhys.85.414.2017.Pantaleone}.
But such experiments are misleading and the small velocity of the chain (or rope), which leads to the failure of the formation of the chain fountain are due to dissipation phenomena which are not taken into account in the theoretical analysis.

In the approach of Refs.~\cite{ProcRSocA.470.20130689.2014.Biggins,EPL.106.44001.2014.Biggins}, an apparent application of the momentum conservation law led to the wrong conclusion that half of the mechanical work done by the tension in the chain may be wasted in the absence of a reaction from the bottom of the container.
Because of this waste, the velocity acquired by the chain in the process of falling over the rim of the beaker is not enough to produce the fountain effect.
In this situation, the solution proposed by the authors of Refs.~\cite{ProcRSocA.470.20130689.2014.Biggins,EPL.106.44001.2014.Biggins} and adopted since then (with one exception known by us~\cite{arXiv1810.13008.Yokoyama}) was that the chain beads receive extra momenta by being kicked off from the bottom of the container.
This additional momentum makes it jump high into the air and siphon from the beaker.

We argue here that the kicks received by the beads from the bottom of the container are not the main reason (assuming they have \textit{any} relevance) for the chain fountain formation (the same conclusion was reached also in~\cite{arXiv1810.13008.Yokoyama}).
If energy and momentum conservation are properly taken into account (while the dissipation effects are correctly evaluated), then the chain may gain enough speed to siphon from the container, without relying on ``kicks''.
Due to its own inertia, the chain cannot simply change from moving upwards to moving downwards at the rim of the beaker, but forms an arch up in the air--like going over an imaginary pulley.
Such a phenomenon can be observed in other activities, like when a rope or a chain is pulled with high velocity over a pulley or an obstacle.
If the tension is not strong enough, the chain (or rope) simply detaches from the pulley and turns at a certain distance from it, as if a second, imaginary pulley is formed.
We support our theoretical model also by some brief experimental observations.
For example, in~\cite{FountainChain.2020.Anghel} we show that a chain placed in a single layer on a soft mat still forms a prominent fountain, although not only the kicks from the bottom are eliminated, but the mat rather increases the dissipation.
Other, unpublished works--where the kicks were eliminated by hanging the chain--support our conclusions~\cite{YouTube.2019.02.10.Yokoyama,YouTube.2019.02.09.Yokoyama}.

The paper is organized as follows.
In the next section, we describe the chain and write the main equations governing its dynamics.
These equations are generally known, but we include them here to make the paper self-contained and to put them in a convenient form.
We write the differential equations that describe the stationary regime and the fluctuations around it. 
Using these equations, we find  the shape  of the  chain in the stationary case and show that it is unstable. 
Eventually, the stationary solution may represent the average over many experimental realizations, but we did not go into details with this analysis.
In Section~\ref{sec_en_mom_cons} we analyze the energy and momentum conservation in the chain, in general, and in particular during the pickup process.
We show that both, energy and momentum, are conserved when there is no dissipation and therefore the velocity acquired by the chain in the pickup process is sufficient to produce the chain fountain.
In our model, the chain is (perfectly) flexible (i.e., there is no rigidity or minimum curvature radius), so there cannot be any ``kicks''.
Therefore, we clearly invalidate the ``scientific consensus''.
Our theoretical analysis is supported by various experiments, in which the ``kicks'' are eliminated by different methods.
In Section~\ref{sec_conclusions} we draw the conclusions.

\section{Mathematical description} \label{sec_chain}

\begin{figure}[t]
  \centering
  \includegraphics[width=6cm,keepaspectratio=true]{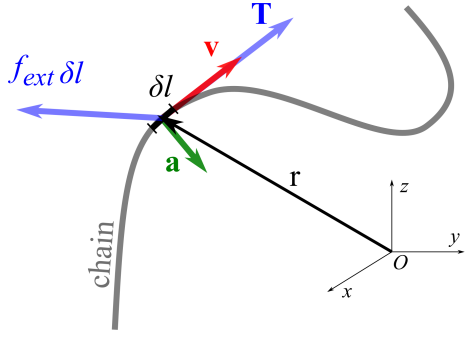}
  \caption{The chain in stationary conditions.
  A chain element $\delta l$, located at $\br[x(l), y(l), z(l)]$, has the velocity $\bv$ and acceleration $\ba$, such that $\bv \cdot \ba = 0$.
  The forces that act on this chain element are the tension $\bT$ and the external force $\delta F_{ext} \equiv f_{ext} \delta l$, where $f_{ext}$ is the external force density.}
  \label{chain_general}
\end{figure}

A non-extensible chain, like the one presented in Fig.~\ref{chain_general}, of constant linear density $\lambda$, is described by the position vector $\br(l, t) \equiv [x(l, t), y(l, t), z(l,t)]$, which depends on two parameters: the position along the path of the chain $l$ and time $t$.
The local velocity (with respect to the laboratory frame) and acceleration are $\bv(l, t) \equiv [v_x(l, t), v_y(l, t), v_z(l,t)]$ and $\ba(l, t) \equiv [a_x(l, t), a_y(l, t), a_z(l,t)] = d\bv(l,t) / dt$, respectively.

\subsection{The stationary case and fluctuations} \label{subsec_descr_chain}

\begin{figure}[t]
  \centering
  \includegraphics[width=5cm,keepaspectratio=true]{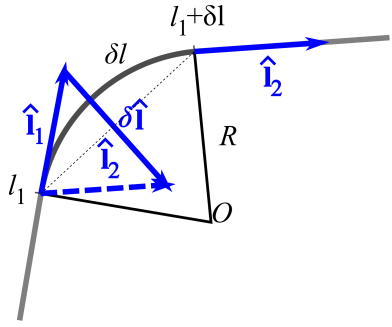}
  \caption{A detail from the chain trajectory.
  Taking the limit $\delta l \to 0$ and using the fact that $|\hat\bl_1| = |\hat\bl_2| = 1$, one can easily prove that $d\hat\bl/dl \equiv c\bl_\perp = \bl_\perp/R$, where $R$ is the curvature radius.}
  \label{deriv_Radius}
\end{figure}

In the stationary case, the chain is moving along a path which does not depend on $t$, $\br(l) \equiv [x(l), y(l), z(l)]$.
The chain is not extensible, so its velocity $\bv(l) \equiv \bv[x(l), y(l), z(l)]$ has the same modulus and is tangent to the path in all its points.
The differential of $\br$ is
\begin{eqnarray}
  && d \br = \left( \hat\bx \frac{dx}{dl} + \hat\by \frac{dy}{dl} + \hat\bz \frac{dz}{dl} \right) dl , \quad {\rm where} \label{def_dr} \\
  && dr = dl \quad {\rm and}
  \quad \left( \frac{dx}{dl} \right)^2 + \left(\frac{dy}{dl}\right)^2 + \left(\frac{dz}{dl}\right)^2 = 1 . \nonumber
\end{eqnarray}
We denote $\hat\bl \equiv \hat\bx \frac{dx}{dl} + \hat\by \frac{dy}{dl} + \hat\bz \frac{dz}{dl}$, where $|\hat\bl| \equiv \hat l = 1$.
From~(\ref{def_dr}) we get
\begin{subequations} \label{def_va}
\begin{eqnarray}
  \bv(l) &\equiv& \frac{d\br(l)}{dt} = \hat\bl(l) v
  \qquad {\rm and} \label{def_v} \\
  \ba (l) &\equiv& \frac{d\bv}{dt} = \left( \hat\bx \frac{d^2x}{dl^2} + \hat\by \frac{d^2y}{dl^2} + \hat\bz \frac{d^2z}{dl^2} \right) v^2 . \label{def_a}
\end{eqnarray}
\end{subequations}
From the invariance of $\delta l$ we obtain $\bv \cdot \ba = 0$--that is, the acceleration is always perpendicular on the local velocity and, therefore, on the path.
We introduce the notations $c(l) \hat\bl_\perp(l) \equiv \hat\bx \frac{d^2x}{dl^2} + \hat\by \frac{d^2y}{dl^2} + \hat\bz \frac{d^2z}{dl^2}$, where $|\hat\bl_\perp(l)| = 1$, $c(l) \ge 0$, and $\hat\bl_\perp(l) \cdot \hat\bl(l) = 0$ for any $l$.
Using Fig.~\ref{deriv_Radius}, one can see that $c(l) \hat\bl_\perp(l) = d\hat\bl/dl = \bl_\perp(l)/R$, where $R$ is the  radius of the curvature of the chain trajectory at $l$.

\begin{figure}[t]
  \centering
  \includegraphics[width=5cm]{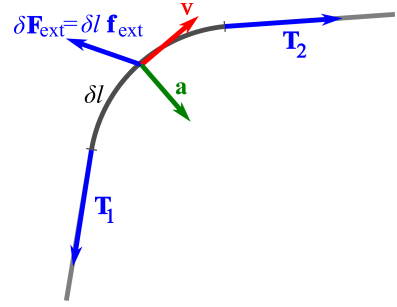}
  \caption{The forces that act on a small chain element $\delta l$ and the acceleration produced. $\bT_1$ and $\bT_2$ are the tensions that act on the element's ends; $\delta \bF_{ext}$ is the external force caused by the force density $\ff_{ext}$--e.g. gravitational force, but the orientation here is arbitrary, for the sake of generalization.}
  \label{F_cp_a_cp}
\end{figure}

The forces that act on the chain are the tension $\bT(l) \equiv T(l) \hat\bl(l)$ (we assume that the chain is not stiff, so $\bT$ acts along the chain) and the external force density $\ff_{ext}(l)$.
The resultant force that acts on the chain element $\delta l$ is (see Fig.~\ref{F_cp_a_cp})
\begin{equation}
  \delta \bF = \delta l \left(\ff_{ext} + \frac{dT}{dl} \hat\bl
  + T c \hat\bl_\perp \right) ,
  \label{Force_tot}
\end{equation}
so we can write Newton's law
\begin{equation} \label{Newton_law}
    \delta m \ba = \delta l \lambda \ba = \delta l \left(\ff_{ext} + \frac{dT}{dl} \hat\bl + T c \hat\bl_\perp \right) .
\end{equation}
%
From Eqs.~(\ref{def_va})-(\ref{Newton_law}) we obtain in the stationary case
\begin{eqnarray}
  && \lambda v^2 c \hat\bl_\perp = \ff_{ext} + \frac{dT}{dl} \hat\bl + T c \hat\bl_\perp . \label{eq_acceleration}
\end{eqnarray}
Let us introduce a local right handed system of coordinates $(\hat\bl, \hat\bl_\perp, \hat\bl_t)$, where $\hat\bl \cdot \hat\bl_t = \hat\bl \cdot \hat\bl_\perp = \hat\bl_\perp \cdot \hat\bl_t = 0$.
Then, we may write $\ff_{ext} \equiv \ff_\perp +  \ff_t + \ff_\parallel \equiv \hat\bl_\perp f_\perp + \hat\bl_t f_t + \hat\bl f_\parallel$, to obtain the conditions for equilibrium,
\begin{subequations} \label{f_equil_eom}
\begin{eqnarray}
  f_\parallel (l) &=& - \frac{dT(l)}{dl} ,  \label{f_par_eom} \\
  f_\perp (l) &=& [\lambda v^2 - T(l)] c(l) , \quad {\rm and} \label{f_perp_eom} \\
  f_t (l) &=& 0 . \label{f_t_eom}
\end{eqnarray}
\end{subequations}
From Eq.~(\ref{f_par_eom}) we see that the variation of $T$ along the chain is only caused by the component of the external force parallel to the chain and compensates it.

Equation~(\ref{f_perp_eom}) is more interesting and we analyse it with the help of Fig.~\ref{F_cp_a_cp}.
First, we observe that if $f_\perp (l) = 0$, then either $\lambda v^2 - T(l) = 0$, or $c(l) = 0$.
If $c(l) \ne 0$ and $\lambda v^2 - T(l) = 0$, the chain may have any curvature radius $R = 1/c(l)$ and the chain trajectory is in indifferent equilibrium--that is, the centripetal force, produced by $T$, equilibrates the chain on any path.


If $c = 0$ and $\lambda v^2 - T(l) \ne 0$, then we have two situations: (1) $T < \lambda v^2$ and (2) $T > \lambda v^2$ (the case $T = \lambda v^2$ was discussed above).
In the case (1), the system is in unstable equilibrium.
For any $c \ne 0$, the centripetal force produced by $T$ is too small to constrain the chain to move on the curved path and therefore it will be displaced even further, changing trajectory.
Therefore, any small disturbance, from $c(l) = 0$ to $c(l) \ne 0$, tends to be amplified.
On the other hand, in the case (2), the system is in stable equilibrium because the centripetal force is too strong and tends to reduce any disturbance, with $c(l) \ne 0$, back to $c(l)=0$.

All the discussions of Eq.~(\ref{f_perp_eom}) for $f_\perp = 0$ applies also to Eq.~(\ref{f_t_eom}), since $f_t(l) = 0$ for any $l$, by definition.

\begin{figure}
    \centering
    \includegraphics[width=4cm]{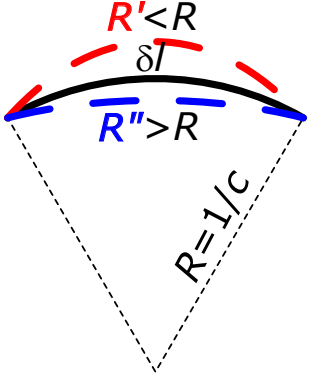}
    \caption{The chain element $\delta l$ with the curvatures $R$ (black, continuous line), corresponding to the stationary trajectory, $R'< R$ (red, dashed line),  and $R'' > R$ (blue, dashed line).}
    \label{fig_curves}
\end{figure}

We continue the discussion of Eq.~(\ref{f_perp_eom}) for the case $f_\perp \ne 0$.
We take $c(l) \ge 0$ by definition, so $f_\perp(l) > 0$ if it is oriented inwards, that is, $\ff \cdot \bl_\perp > 0$.
In such a case, from Eq.~(\ref{f_perp_eom}) we obtain $\lambda v^2 - T(l)>0$.
If the equilibrium is disturbed by increasing $c(l)$ to $c'(l) > c(l)$ (see Fig.~\ref{fig_curves}), then, from Eq.~(\ref{f_perp_eom}) we obtain $f_\perp (l) + T(l) c'(l) < \lambda v^2 c'(l)$.
This means that the total centripetal force $f_\perp (l) + T(l) c'(l)$ is to small to constrain the chain to move along the trajectory with the curvature radius $R' = 1/c' < 1/c$, so the chain will be displaced outwards, in the direction $-\hat\bl_\perp$.
Similarly, if $c(l)$ is changed to $c''(l) < c(l)$ (see Fig.~\ref{fig_curves}), we obtain a centripetal force $f_\perp (l) + T(l) c''(l) > \lambda v^2 c''(l)$ which is too big and therefore displaces the chain trajectory of curvature radius $R'' = 1/c''$ in the direction $\hat\ff_\perp$, that is, inwards.
Therefore, in the situation $f_\perp(l) > 0$, the chain's stationary path is unstable with respect to local deformations which change $c(l)$--if $f_\perp(l)$ does not change with the deformation.
Therefore, as we can see in Fig.~\ref{fig_curves}, if $c(l)$ increases by moving the point $l$ outwards, the inertia overcomes the centripetal force and the chain has the tendency of deviating more, if no other constraints forbid it.
Vice-versa, if $c(l)$ decreases by displacing the chain inwards, the displacement is further amplified because in this case the centripetal force becomes bigger than the $\lambda v^2 c(l)$.
Therefore, in the case $f_\perp(l) > 0$ the stationary trajectory may not be stable, since the fluctuations tend to be amplified.

The situation is opposite if $f_\perp(l) < 0$.
Then, $\lambda v^2 - T(l)>0$ and if $c(l)$ is changed into $c'(l) > c(l)$, we obtain $f_\perp + T c' > \lambda v^2 c'$, which means that the centripetal force overcomes the inertia of the chain and pulls it inwards, in the direction of $\bl_\perp$.
On the other hand, if $c(l)$ changes into $c''(l) < c(l)$, we obtain $f_\perp + T c'' < \lambda v^2 c''$ which means that the chain moves outwards, in the direction $-\bl_\perp$.
Therefore, we may conclude that the stationary solution is stable when $f_\perp(l) < 0$.

\subsection{Solutions for the stationary shape of the chain} \label{subsec_sol_stat}

\begin{figure}[b]
  \centering
  \includegraphics[width=4cm,keepaspectratio=true]{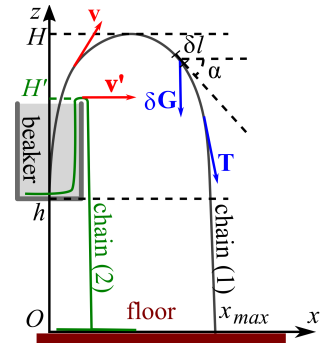}
  \caption{The schematic representation of a chain falling from the beaker onto the floor in the cases when the chain fountain is formed (1) and when it is not formed (2).
  The bottom of the beaker is at height $h$ above the floor and the chain is moving with velocity $\bv$~(1) or $\bv'$~(2).
  The gravity force that acts on the chain segment $\delta l$ is $\delta \bG = - \lambda g \delta l \hat \bz$ and $\hat \bl \cdot \hat \bx = \cos(\alpha)$.
  }
  \label{chain_fall}
\end{figure}

The stationary shape of the chain was found before (see, for example, \cite{EPL.106.44001.2014.Biggins}) and may be obtained from Eqs.~(\ref{f_equil_eom}), with proper boundary conditions.
The situation is schematically illustrated in Fig.~\ref{chain_fall}, where we represent a chain falling from a beaker onto the floor.
The beaker is at height $h$ above the floor and the force that acts on the chain is the gravity, $\ff_{ext} = \lambda \bg$, where $\bg = -g \hat\bz$ is the gravitational acceleration, acting along the $z$ direction, downwards.
Due to the symmetry of the problem, we can choose the coordinate axes $(x,y,z)$ such that $y\equiv 0$ -- that is, we work in the $(x,z)$ plane.
The parametric curve $[x(l), z(l)]$, that describes the chain in this plane will be changed into the function $z(x)$, which is single valued and defined in the interval $[0, x_{max}]$.
From Eqs~(\ref{def_dr}) and (\ref{def_va}), together with (see Fig.~\ref{chain_fall}) $dz/dx \equiv z' = (dz/dl)/(dx/dl) = \tan(\alpha)$, we obtain
\begin{eqnarray}
  && \left| \frac{dx}{dl} \right| = \frac{1}{\sqrt{1+(z')^2}}, \quad
  \left| \frac{dz}{dl} \right| = \frac{|z'|}{\sqrt{1 + (z')^2}} , \quad
  \nonumber \\
  && \left| \frac{d^2z}{dl^2} \right| = \frac{|z''|}{[1+(z')^2]^{2}} , \quad
  %
  \left| \frac{d^2x}{dl^2} \right| = \frac{|z' z''|}{[1+(z')^2]^{2}} ,
  \label{derivs_xz} \\
  && {\rm and} \quad
  c = \frac{|z''|}{[1+(z')^2]^{3/2}} \equiv \frac{1}{R} .
  \nonumber
\end{eqnarray}
From Eq.~(\ref{f_par_eom}) we obtain $f_\parallel = g \lambda \sin(\alpha) = - T' (dx/dl) = -T' \cos(\alpha)$, where $T' \equiv dT/dx$.
Taking into account that $z' = \tan(\alpha)$, we further obtain
\begin{equation}
  T' = - g\lambda\tan(\alpha) = - g\lambda z', \quad
  T[z(x)] = g\lambda z + T_0 , \label{val_T_z}
\end{equation}
where $T_0$ is the tension in the chain at $z=0$.
Assuming that the tension  becomes zero when the chain reaches the floor ($z=0$), in the rest of the calculations we shall assume $T_0 = 0$~\cite{ProcRSocA.470.20130689.2014.Biggins,EPL.106.44001.2014.Biggins,FrontiersinPhysics.6.84.2018.Flekkoy}.
If $T_0 \ne 0$, one can adjust the position $z=0$ below or above the floor, such that the condition $T(z=0) \equiv T_0 = 0$ is still satisfied.

From Eq.~(\ref{f_perp_eom}) and using $f_\perp = g\lambda \cos\alpha = g\lambda / \sqrt{1 + z'^2}$ we obtain
\begin{eqnarray}
  g \lambda &=& [\lambda v^2 - T(x)] \frac{|z''|}{1+(z')^2} . \label{eq_traj}
\end{eqnarray}
We observe that $\lambda v^2 - T(x) > 0$ and $f_\perp > 0$ along the whole trajectory, so, according to the discussion in Section~\ref{subsec_descr_chain}, the stationary trajectory is not a stable path.

Taking into account that along the chain trajectory $z'' < 0$ and replacing $T[z(x)]$ from Eqs.~(\ref{val_T_z}) and (\ref{eq_traj}) we get
\begin{eqnarray}
  g &=& - [ v^2 - g z ] \frac{z''}{1+(z')^2} . \label{eq_traj2}
\end{eqnarray}
We see that the stationary trajectory does not depend on $\lambda$.
Furthermore, if we replace the coordinates $(x,z)$ by the dimensionless coordinates $(x_1, z_1)$, where
\begin{equation}
  x_1 \equiv \frac{gx}{v^2} \quad {\rm and} \quad z_1 \equiv \frac{gz}{v^2} , \label{defs_x1_z1}
\end{equation}
Eq.~(\ref{eq_traj2}) simplifies to
\begin{equation}
  z_1'' (z_1-1) - (z_1')^2 -  1 = 0 , \label{main_eq_z}
\end{equation}
with the solution
\begin{equation}
  z_1 (x_1) = C_1 \cosh \left( \frac{x_1 + C_2}{C_1} \right) + 1 , \label{sol_eq_z1}
\end{equation}
where $C_1$ and $C_2$ are two constants that have to be determined from the boundary conditions.

\section{Energy and momentum conservation and the formation of the chain fountain} \label{sec_en_mom_cons}

From Eq.~(\ref{f_perp_eom}) we observe that under gravity (or any external force), the chain bends with finite radius ($c>0$) in the direction of the force ($f_\perp \ge 0$) if and only if
\begin{equation}
    \lambda v^2 > T . \label{cond_fountain}
\end{equation}
%
If, for example, $\lambda v^2 \to T$, then the condition~(\ref{f_perp_eom}) is satisfied if $c \to \infty$ ($R \to 0$), so the chain may change direction sharply, under the influence of any--arbitrarily small--external force.
If this happens at the rim of the beaker, then the chain fountain cannot be formed.
So, the necessary condition to have a chain fountain is that $\lambda v^2 > T(H')$ (see Fig.~\ref{chain_fall}).
On the other hand, in the previous section we showed that for the stationary trajectory, $T$ increases with the height $z$ (Eq.~\ref{val_T_z}), so $T(H') > T(h)$.
Therefore, the minimal condition to have the chain fountain is that $\lambda v^2 > T(h)$.

To see what velocity an ideal chain may acquire when it is picked up from the bottom of a container, let us use an example.

\begin{figure}[t]
    \centering
    \includegraphics[width=6.5cm,keepaspectratio=true]{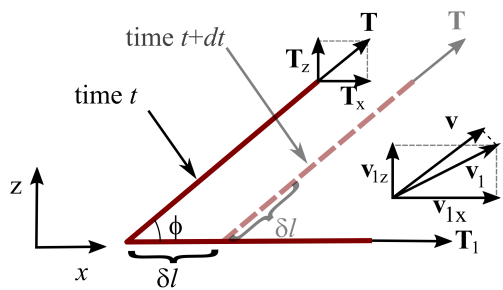}
    \caption{A very simple, 2D model, in  which a very short part of an ideal chain, of mass density $\lambda$, it is pulled up by the tension force $\bT$. The part of the chain along the $x$ direction is supposed to be kept immobile by a resultant force $\bT_1$ (for example, it is attached to the rest of the chain and the friction between the chain and the substrate is strong enough to keep it immobile). The moving part of the chain has the speed $v_1$ and the chain's shapes at time $t$ and $t+\delta t$ are represented by solid and dashed brown lines, respectively.
    The moving part of the chain makes an angle $\phi$ with the $x$ axis and it is parallel to $\bv$ and $\bT$.
    The gravity is not represented, being irrelevant, and therefore the 2D figure may be turned in any position. In one particular case (when the chain is hanging), the figure is turned in such a way that $x$ is vertical and $y$ is horizontal.
    }
    \label{pulling_ideal_chain}
\end{figure}

First, we notice that Eqs.~(\ref{Force_tot}) and (\ref{Newton_law}) determine the general movement of the chain and they conserve both, energy (the mechanical work done by the forces $T$ and $f_{ext}$ is regained as kinetic energy of the chain) and momentum (explicitly, Eq.~\ref{Newton_law}), since no explicit dissipation mechanism was taken into account.
The same happens in the pickup process.

Let's analyze the movement of the ideal chain from Fig.~\ref{pulling_ideal_chain}.
The example is two-dimensional (2D) and the gravity plays no role, the length of the chain considered being very small.
Since the gravity is irrelevant, the arguments below are valid no matter how we turn the 2D figure with respect to the vertical direction.
In general, $x$ is horizontal and $y$ is vertical, but in one particular case (the hanging chain), $x$ is vertical and $y$ is horizontal.

We assume that the oblique part of the chain in Fig.~\ref{pulling_ideal_chain}, which is pulled by the tension $\bT$, moves at constant speed $\bv_1$ ($\bv_1$ is not parallel to $\bT$), whereas the horizontal part is at rest, under the tension $\bT_1$.
The part of the chain which is at rest makes an angle $\phi$ with the part that is moving.
$\bT$ is parallel to the moving part of the chain, $v_{1x}$ and $v_{1z}$ are the components of $v_1$ along the $x$ and $z$ axis, and $\bv$ is the projection of $\bv_1$ along the direction of $\bT$.
Then, geometrical considerations lead to 
\begin{equation}
\frac{v_{1x}}{v_{1z}} 
= \frac{1 + \cos\phi}{\sin\phi} , \ 
v_{1x} = v, \ 
{\rm and} \ 
v_1^2 = \frac{2 v^2}{1+\cos\phi} . \label{vel_rels}
\end{equation}
Equations~(\ref{vel_rels}) and the momentum conservation along $x$ and $z$ directions give
\begin{equation} \label{Eqs_mom_cons}
    T \cos\phi + T_1 = \frac{ \lambda v^2 }{1 + \cos \phi}
    \ {\rm and} \ 
    T = \frac{\lambda v^2}{(1 + \cos \phi)^2} ,
\end{equation}
respectively.
From Eqs.~(\ref{Eqs_mom_cons}) we get $T_1 = T$ and combining Eqs.~(\ref{vel_rels}) and (\ref{Eqs_mom_cons}) we obtain
\begin{equation}
T = \frac{\lambda v_1^2}{2(1+\cos \phi)}
\equiv \frac{\epsilon_k}{1+\cos \phi}
, \label{Eq_T}
\end{equation}
where $\epsilon_k$ is the linear density of kinetic energy of the moving chain.

The energy conservation may be easily checked.
The mechanical work produced by the traction force $\bT$ is $L = T v \delta t = T \delta l (1 + \cos \phi)$, whereas the kinetic energy gained by the segment $\delta l$ is $E_k = \delta l \lambda v_1^2 / 2$. Equating $L$ and $E_k$ gives Eq.~(\ref{Eq_T}), which certifies the consistency of our formalism and the conservation of both, energy and momentum.

Let us now look at some specific cases.
For $\phi = \pi/2$ (vertical pickup of a horizontal chain) we have $T = \epsilon_k = \lambda v_1^2/2 = \lambda v^2$, whereas for $\phi = 0$ (vertical pickup of a hanging chain) we have $T = \epsilon_k/2 = \lambda v_1^2/4 = \lambda v^2/4$.
In both these cases, $\lambda v_1^2 > T$, which implies that the chain fountain can be formed, contrary to the ``scientific consensus'', which states that momentum conservation during the pickup process leads to the dissipation of half of the energy, so the chain fountain cannot happen unless the beads of the chain are kicked off the ground or kicked off each-other~\cite{ProcRSocA.470.20130689.2014.Biggins,EPL.106.44001.2014.Biggins,PhysEduc.50.564.2015.Andrew,ProcRSocA.471.20140657.2015.Virga,AmJPhys.85.414.2017.Pantaleone,ArchApplMech.87.1411.2017.Pfeiffer,FrontiersinPhysics.6.84.2018.Flekkoy,FrontiersinPhysics.7.187.2019.Flekkoy,Wikipedia.chain.fountain}.
Furthermore, the irrelevance of the ``kicks'' for the formation of the chain fountain was demonstrated also experimentally in~\cite{FountainChain.2020.Anghel}.
Moreover, in the examples given here there are no beads.

\begin{figure}[t]
    \centering
    \includegraphics[width=2 cm,keepaspectratio=true]{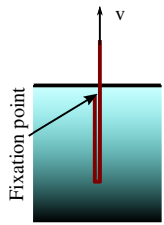}
    \caption{The configuration that may lead to maximum fountain effect: a vertically hanging chain is pulled (also vertically) over an obstacle.}
    \label{vertical_chain}
\end{figure}

From Eq.~(\ref{Eq_T}) we can find that the condition for the formation of the chain fountain, namely $\lambda v_1^2 > T$, which is satisfied for any $0  \le \phi < \arccos(-0.5) = 2\pi/3$.
In the other extreme, the configuration most favorable for the formation of the chain fountain is the one with $\phi = 0$,  in which the chain is hanging from an obstacle and siphoning vertically over it, as schematically shown in Fig.~\ref{vertical_chain}.
For the same tension in the chain, this configuration produces maximum velocity and therefore highest fountain, plus it eliminates the possibility of ``kicks'' from the bottom of the container.
This phenomenon was nicely demonstrated in~\cite{YouTube.2019.02.10.Yokoyama,YouTube.2019.02.09.Yokoyama}.

\section{Conclusions} \label{sec_conclusions}

We analyzed the dynamics and the stationary trajectory of the chain fountain produced by an ideal, (perfectly) flexible chain--that is, the infinitely thin chain, with no beads (or links) and no internal dissipation, has a mass density $\lambda$, no lateral rigidity and may turn at any angle.
Special attention was devoted to the pickup process and it was shown that both, energy and momentum are conserved (that is, the mechanical work and the momentum produced by the tension in the chain and the external forces are recovered as kinetic energy and momentum of the chain set in motion), as one should expect from a process with no dissipation.
Because of this, the velocity acquired by the chain in the pickup process is enough to produce a chain fountain effect and there is no need for ``kicks'' from the bottom of the container.
In this way, we invalidate the (so called) ``scientific consensus'', that no chain fountain is possible for a flexible chain, and the only mechanism capable to give the chain enough velocity to siphon from the container must be the ``kicks'' from the supporting surface.
Whereas a small upwards reaction from the supporting surface is possible due to the lateral rigidity and the momentum of inertia of the beads or links of a real chain, such a reaction is not necessary for the chain fountain effect and, in most cases, at best, should have negligible effects due to the small momentum of inertia of the beads.
Moreover, when beads come in contact with each-other, the random collisions between them should rather be a cause for dissipation, not for a coherent upwards movement, as it is supposed in the ``scientific consensus''.
To experimentally invalidate the effect of the ``kicks'' on the chain fountain formation, we realized a short video in which the chain is laid in a single layer on a soft mat~\cite{FountainChain.2020.Anghel}.
Although the mat would eventually only absorb energy, we can still see a very strong chain fountain.

Our results regarding velocity, momentum and energy calculations differ from the ``scientific consensus'' because we took into account also the forces that act on the part of the chain (and \textit{in the chain}) which is not picked up (see Fig.~\ref{pulling_ideal_chain}).
These forces should be composed with the ones acting on the moving part of the chain to correctly calculate the velocities.
In this way, the formalism becomes consistent and the chain fountain effect possible, even in flexible, infinitely thin (but with finite linear mass density) chains.

We theoretically showed what are the conditions (e.g. pickup angles) for the formation of the chain fountain and which is the configuration that may produced the most pronounced effect (see Fig.~\ref{vertical_chain}).
The most favorable configurations are the ones in which the chain is hanging and is lifted vertically.
In such configurations there is no substrate to kick-off the beads and our model is clearly tested.
They have been realized by Prof. Yokoyama~\cite{YouTube.2019.02.09.Yokoyama,YouTube.2019.02.10.Yokoyama} and confirm the theoretical model.

We also observed that the stationary trajectory is unstable.
The stability of the trajectory may be increased by the lateral stiffness of the chain, but a more detailed analysis of the fluctuations and of the averages over many realizations is necessary.

\begin{acknowledgements}
I am grateful to Prof. Eirik Grude Flekk{\o}y for showing me the chain fountain.
Similar conclusions as mine have been independently obtained by Prof. Hiroshi Yokoyama, who noticed my manuscript and kindly draw my attention to his works~\cite{arXiv1810.13008.Yokoyama,YouTube.2019.02.09.Yokoyama,YouTube.2019.02.10.Yokoyama}.

The work was supported by the UEFISCDI project PN 19060101 / 2019, ELI-RO contract 81-44 /2020, and Romania-JINR collaboration projects.
\end{acknowledgements}

\section*{Declarations}

Funding: Unitatea Executiv\u{a} pentru Finan\c{t}area \^{I}nv\u{a}\c{t}\u{a}m\^{a}ntului Superior, a Cercet\u{a}rii, Dezvolt\u{a}rii \c{s}i Inov\u{a}rii (UEFISCDI) PN 19060101 / 2019, ELI-RO contract 81-44 /2020, and Romania-JINR collaboration projects\\
Conflict of Interest: The authors declare that they have no conflict of interest.\\
Availability of data and material (data transparency): not applicable.\\
Code availability (software application or custom code): not applicable.\\
Authors' contributions: not applicable.



\end{document}